\documentclass{iaus}
\usepackage{graphicx}

\title[Dynamical effects of the long bar in the Milky Way] 
{Dynamical effects of the long bar in the Milky Way}

\author[Gardner et al.]{Esko Gardner$^1$, Kimmo A. Innanen$^2$ \and Chris
  Flynn$^1$}

\affiliation{$^1$Tuorla Observatory, Department of Physics and
  Astronomy, University of Turku \\
  V\"aisal\"antie 20, FI-21500 Piikki\"o, Finland\\
e-mail: esgard@utu.fi\\[\affilskip]
$^2$Dept. of Physics and Astronomy\\ York University, Toronto, Canada
}

\pubyear{2008}
\volume{254}  
\jname{The Galaxy Disk in Cosmological Context}
\editors{J. Andersen (Chief Editor), J. Bland-Hawthorn \& B. Nordstr\"om}
\begin{document}

\maketitle

\begin{abstract} 

We examine the dynamical effects on disk stars of a ``long bar'' in the Milky
Way by inserting a triaxial rotating bar into an axisymmetric
disk$+$bulge$+$dark halo potential and integrating 3-D orbits of 10$^4$ tracer
stars over a period of 2 Gyr. The long bar has been detected via ``clump
giants'' in the IR by \cite{LC07}, and is estimated to have semi-major axes of
($3.9\times0.6\times0.1$) kpc and a mass of $6\times10^9$ M$_\odot$. We find
such a structure has a slight impact on the inner disk-system, moving
tracers near to the bar into the bar-region, as well as into the bulge. These effects are under
continuing study.

\end{abstract}

\firstsection 
              
\section{Introduction}

Our initial interest was to study a barred galaxy potential, and especially the
resonance effects which might lead to vertical ejection into the halo of
globular clusters formed in the disk as suggested by \cite{Innanen07}. We took
a triaxial ellipsoid, a Ferrers' potential, to model the bar (see
\cite{Pfenniger84} for a similar application of the potential as a bar). We
decided to look at the effects of the ``long bar'' described by \cite{LC07},
taking their parameters directly, and applying it to our system. The dynamical
effects of the bar are quite interesting, and it should be possible to
constrain the bar parameters with analysis of these effects.

We also present a brief discussion on the scale-length of the Galactic disk,
and a revised version of the \cite{Flynn96} Galactic disk model with a shorter
scale-length of 3 kpc rather than 4.4 kpc, to be more consistent with recent
observations.

\section{Scale-length of the Milky Way disk}

Our motivation in constructing a disk model with a revised disc scale-length
stems from recent infrared observations, which show a considerably shorter
scale-length for the Galactic disk than optical observations. The disk
scale-length ($h_R$) adopted in \cite{Flynn96} was 4.4 kpc, based on the
kinematic determination (using velocity dispersions of old disk red giants) by
\cite{Lewis89}, who obtained a scale-length of $h_R = 4.4 \pm 0.3$ kpc. It has
been long known that the scale-length measured in the infrared is considerably
shorter than in the optical, Pioneer 10 data giving $h_R = 5.5 \pm 1.0$ kpc
already in the 1980s (\cite{Pioneer10}). More recently near-IR data in J and K
with DENIS give $h_R = 2.3 \pm 0.1$ kpc and 2MASS of K giants gives $h_R = 3.34
\pm 0.1$ kpc (\cite{LC02,Ruphy96,Ojha01}). These latter surveys probe the
scale-length of emitted light, mainly by giant stars, rather than probing
directly the scale-length of the mass, which is what we are actually interested
for modeling the disk's potential. M dwarfs do trace the mass directly:
star-counts with HST in the 1990s of disk M dwarfs indicate a disk scale-length
of $3.0 \pm 0.4$ kpc (\cite{Gould97}). On the other hand, local kinematics of
disk stars, which should also be mostly sensitive to the mass rather than the
light distribution, give a range of possible scale-lengths 1.7$-$2.9 kpc
(\cite{Bienayme97}). Very recently, F to K type dwarfs, for which distances and
metallicities have been determined in the huge numbers in 6500 square degrees
of sky and probing to distances of up to 20 kpc above and below the disk near
the Sun using SDSS data, give a scale-length of $2.6\pm 0.5$ kpc
(\cite{Juric08}).

\section{The disk model}

\begin{figure}[!htbp]
\begin{center}
 \includegraphics[width=3in]{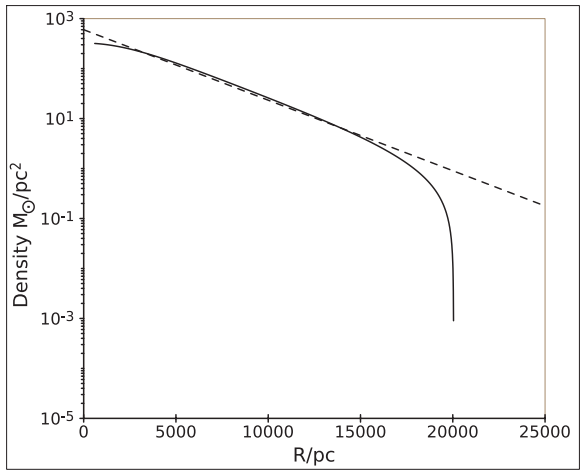} 
 \caption{The surface density of the disk component as a function of
   Galactocentric radius.  The dashed line corresponds to an exponential
   density falloff, 3 kpc, which is a good fit to the model over a wide range
   of radii. Note that the density truncates strongly at 18 kpc.}
   \label{fig1}
\end{center}
\end{figure}

The \cite{Flynn96} model of the Galactic disk has a $h_R = 4.4$ kpc, and is
built from three superimposed Miyamoto-Nagai potentials (with different linear
scales) to obtain the exponentially falling density profile. In light of recent
observations, we would like a shorter disk scale-length, and modified the
\cite{Flynn96} model to a disk with $h_R = 3.0$ kpc. The local volume density
local surface density of the disk are as in the 1996 model (i.e. $\rho_0
\approx 0.1 M_\odot/pc^3$ and $\Sigma_\odot \approx 50 M_\odot/pc^2$), and are
consistent with local measurements (\cite{Holmberg00}). We adopt a solar
Galactocentric distance of R$_\odot = 8$ kpc.  Figure \ref{fig1} shows the
surface density profile of the disk model, with the dashed line indicating a
scale-length of 3 kpc. This is a good fit over a wide range of Galactocentric
radii (2 to 15 kpc). Note that the disk truncates strongly at approximately 18
kpc.

\section{Inserting the long bar}

We insert a triaxial bar potential into the central regions of the model.  We
chose a uniform density Ferrers' potential (see eg. \cite{EFE}, or
\cite{GalDyn}). The linear semi-major axes (3.9:0.6:0.1 kpc), and mass ($6
\times 10^9 M_\odot$) were chosen from \cite{LC07}, who used ``red clump''
stars detected in a 10 degree wide region observed in the near IR along the
Galactic plane. Note that the long bar is very thin vertically, a property
which leads to some interesting effects on orbits of disk stars which come near
it. The bar is set to rotate around the $z$-axis with with a pattern speed of
50 km/s/kpc, giving it a speed of 200 km/s at its tip. This is the speed
suggested by OLR (\cite{Dehnen00}), although we have tried a range of pattern
speeds from 0 to 75 km/s/kpc, without substantially altering the conclusions of
the paper. The initial phase angle of the bar is as observed by \cite{LC07},
i.e. 43 degrees.  An important point to note is that we aim to investigate the
effects of the ``long bar'', rather than the smaller ``COBE bar''
(\cite{Bissantz02}), which has dimensions (3.5:1.4:1) kpc, and a position angle
of 22 degrees is rather dissimilar to the long bar.

 \section{Disk simulation}

We set up an exponential distribution of tracer stars with the same
scale-length as the disk. We chose $U, V$, and $W$ velocities from a normal
distribution with a standard deviation of 10 km/s around the local standard of
rest (i.e.  the circular velocity) in the total potential, so that the stars
are very nearly on exact circular orbits. Thus the disk is initially inherently
cold, and aids in following the secular development of the long bar's kinematic
effects. We integrated the orbits of 10$^4$ objects in steps of 10$^4$ years,
with a total time of 2 Gyr, in both the barred and unbarred systems,
using the long bar with the parameters advocated by \cite{LC07}. We also examined the effects of a thicker, less dense
long bar, with an axial ratio of (3.9:2:1) kpc.

\section{Results and discussion} 

\begin{figure}[!htbp]
\begin{center}
 \includegraphics[width=5in]{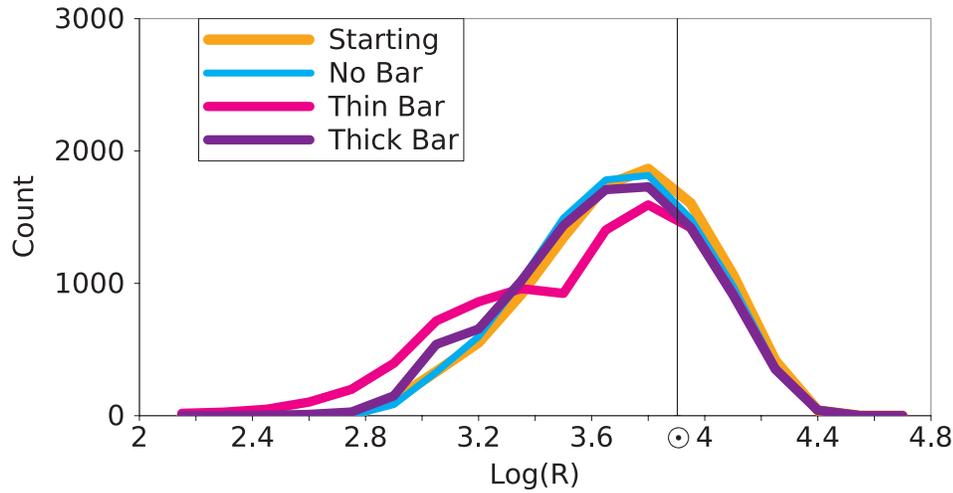} 
 \caption{Histograms of Galactocentric radii of the tracer stars, for the
   initial distribution (orange), and after all three simulations have
   run (2 Gyr). The vertical line marks the position of the Sun at 8
   kpc from the  Galactic center. The ``no bar'' case (cyan) shows no secular evolution,
   indicating the initial setup in the disk is stable. Cases with the bar
   turned on show some secular evolution. Analysis shows that the long
   thin bar (magenta) has slight dynamical effects on disk stars which
   pass near it, injecting the inner disk, and bulge, with stars that have orbits
   near the tip of the bar. The thicker bar exibhits a feature around
   1 kpc.}\label{fig2}
\end{center}
\end{figure}

\begin{figure}[!htbp]
\begin{center}
 \includegraphics[width=5in]{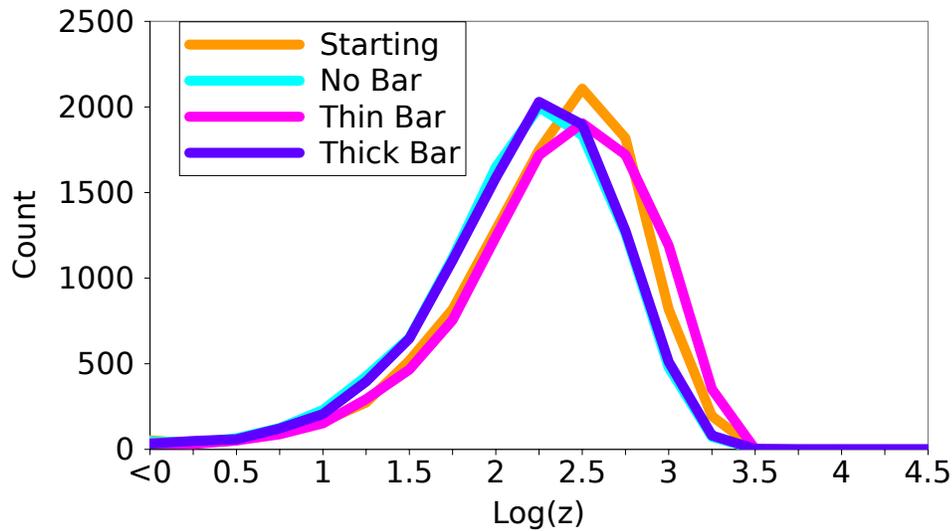}
 \caption{Histograms of the vertical height of the tracer stars,
   initially, and
   at the end of 2 Gyrs integration, for the ``no bar'' case (cyan), the long bar and the
   thicker long bar cases (magenta, blue).  No secular evolution is
   seen in the distributions with the bar, although there is a slight
   trend to higher $z$-values in the thin bar case.}\label{fig3}
\end{center}
\end{figure}

We present here a preliminary look at the results. In Figures \ref{fig2}, and
\ref{fig3}, we show the final distribution of the objects, after 2 Gyr of
orbital integrations, for the cases of ``no bar'', a long bar of high density
(6 $M_\odot/pc^3$) and a long bar with a much lower density (0.2
$M_\odot/pc^3$). The initial (orange), and final radial distributions of the
objects are shown in Figure \ref{fig2}. The vertical distribution is shown in
Figure \ref{fig3}.  In both, the ``no bar'' case (cyan), shows little secular
evolution, indicating that the disk has been set up in an stable manner.  In
the denser long bar (magenta), the bar is injecting the inner regions
with disk stars from approximately 1 kpc beyond the tip of the bar.
Within the bar region, significant modification of the distribution profile
takes place, as one would expect. This is seen as a knee in the
distribution profile at $\log R = 3.0-3.6$. The less dense, thick bar (blue), has much
smaller effects, with a small knee forming around 1 kpc,
which coincides with the spherical area inside the bar.\\

The bars have almost no effect on the $z$-distribution of the disk,
there is a slight shift towards higher values in the thin case
(magenta), these are mostly tracers that orbitally evolve towards the
center of the system, ie. into the bulge. There is no significant
ejection to large heights above the disk.

\vskip 1 truecm

Acknowledgements: We thank Martin L{\'o}pez-Corredoira for his useful comments
on the manuscript. EG acknowledges the financial support of the Finnish
Cultural Foundation, and the Finnish Graduate School in Astronomy and Space
Physics. CF and KI are very grateful to the Academy of Finland for financial
support.

\end{document}